\documentclass[11pt,a4paper]{article}
\usepackage[hyperref]{acl2020}
\usepackage{times}
\usepackage{latexsym}
\usepackage{mathptmx}
\usepackage{amsmath}

\usepackage{enumitem,kantlipsum}
\usepackage{microtype}
\usepackage{booktabs}

\usepackage{graphicx}
\usepackage{rotating}
\usepackage{array}
\newcolumntype{L}[1]{>{\raggedright\let\newline\\\arraybackslash\hspace{0pt}}m{#1}}
\newcolumntype{C}[1]{>{\centering\let\newline\\\arraybackslash\hspace{0pt}}m{#1}}
\newcolumntype{R}[1]{>{\raggedleft\let\newline\\\arraybackslash\hspace{0pt}}m{#1}}
\newcolumntype{M}[1]{>{\centering\arraybackslash}m{#1}}
\newcolumntype{x}[1]{>{\centering\arraybackslash\hspace{0pt}}p{#1}}

\aclfinalcopy 
\def\aclpaperid{3085} 

\newcommand{\kareem}{\textcolor{black}}

\title{Predicting the Topical Stance and Political Leaning of Media using Tweets}

\author{
  Peter Stefanov$^1$, Kareem Darwish$^2$, Atanas Atanasov$^3$, Preslav Nakov$^2$ \\
  $^1$SiteGround Hosting EOOD, Bulgaria\\
  $^2$Qatar Computing Research Institute, HBKU, Doha, Qatar \\
  $^3$Sofia University ``St. Kliment Ohridski'', Sofia, Bulgaria\\  
  \texttt{\{stefanov.peter.ps,atanas.atanasov.sf\}@gmail.com}, \\
  \texttt{\{kdarwish,pnakov\}@hbku.edu.qa} \\}

\date{}

\begin{document}
\maketitle
\begin{abstract}
Discovering the stances of media outlets and influential people on current, debatable topics is important for social statisticians and policy makers. Many supervised solutions exist for determining viewpoints, but manually annotating training data is costly. In this paper, we propose a cascaded method that uses unsupervised learning to ascertain the stance of Twitter users with respect to a polarizing topic by leveraging their retweet behavior; then, it uses supervised learning based on user labels to characterize both the general political leaning of online media and of popular Twitter users, as well as their stance with respect to the target polarizing topic. We evaluate the model by comparing its predictions to gold labels from the Media Bias/Fact Check website, achieving 82.6\% accuracy.
\end{abstract}

\section{Introduction}

Online media and popular Twitter users, which we will collectively refer to as \emph{influencers}, often express overt political leanings, which can be gleaned from their positions on a variety of political and cultural issues. Determining their leaning can be done through the analysis of their writing, which includes the identification of terms that are indicative of stance \cite{groseclose2005measure,gentzkow2011ideological}. Performing such analysis automatically can be done using supervised classification, which in turn would require manually labeled data \cite{groseclose2005measure,gentzkow2011ideological,mohammad2016semeval}. Alternatively, leanings can be inferred based on which people share the content (blogs, tweets, posts, etc.) on social media, as social media users are more likely to share content that originates from sources that generally agree with their positions \cite{an2012visualizing,morgan2013news,ribeiro2018media,wong2013quantifying}. 

\noindent Here, we make use of this observation to characterize influencers, based on the stances of the Twitter users that share their content.  
Ascertaining the stances of users, also known as stance detection, involves identifying the position of a user with respect to a topic, an entity, or a claim \cite{mohammad2016semeval}. For example, on the topic of abortion in USA, the stances of left- vs. right-leaning users would typically be ``pro-choice'' vs. ``pro-life'', respectively.  

\begin{figure}
    \centering
    \includegraphics[width=\linewidth]{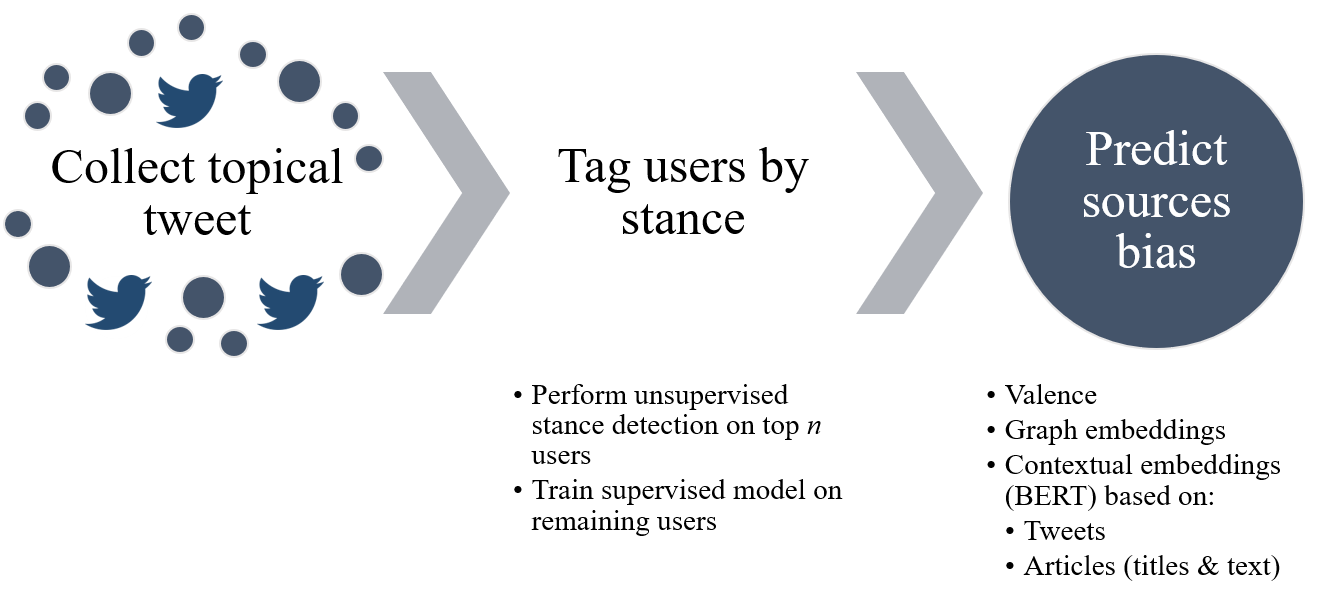}
    \caption{General outline of our methodology.}
    \label{fig:pipeline}
\end{figure}

In this paper, we propose to apply unsupervised stance detection to automatically tag a large number of Twitter users with their positions on specific topics \cite{darwish2019unsupervisedStance}. The tagging identifies clusters of vocal users based on the accounts that they retweet. Although the method we use may yield more than two clusters, we retain the two largest ones, which typically include the overwhelming majority of users, and we ignore the rest. Then, we train a classifier that predicts which cluster a user belongs to, in order to expand our clusters. Once we have increased the number of users in our sets, we determine which sources are most strongly associated with each group based on sharing by each group. We apply this methodology to determine the positions of influencers and of media on eight polarizing topics along with their overall leaning: left, center or right. In doing so, we can also observe the sharing behavior of right- and left-leaning users, and we can correlate their behavior with the credibility of the sources.  Further, given the user stances for these eight topics, we train a supervised classifier to predict the overall bias of sources using a variety of features, including the so-called \emph{valence}~\cite{conover2011political}, graph embeddings, and contextual embeddings.  Using a combination of these features, our classifier is able to predict the bias of sources with 82.6\% accuracy, with valence being the most effective feature.  Figure~\ref{fig:pipeline} outlines our overall methodology.

Our contributions are as follows:
\begin{itemize}
    \item We use unsupervised stance detection to automatically determine the stance of Twitter users with respect to several polarizing topics.
    \item We then use distant supervision based on these discovered user stances to accurately characterize the political leaning of media outlets and of popular Twitter accounts. For classification, we use a combination of source valence, graph embeddings, and contextualized text embeddings.
    \item We evaluate our approach by comparing its bias predictions for a number of news outlets against gold labels from Media Bias/Fact Check. We further evaluate its predictions for popular Twitter users against manual judgments. The experimental results show sizable improvements over using graph embeddings or contextualized text embeddings.
\end{itemize}

The remainder of this paper is organized as follows: Section~\ref{sec:related} discusses related work.  Section~\ref{sec:data} describes the process of data collection.
Section~\ref{sec:method} presents our method for user stance detection.
Section~\ref{sec:influencers} describes how we characterize the influencers.
Section~\ref{sec:media:bias} discusses our experiments in media bias prediction.
Finally, Section~\ref{sec:conclusion} concludes and points to possible directions for future work.

\section{Related Work}
\label{sec:related}

Recent work that attempted to characterize the stance and the ideological leaning of media and Twitter users relied on the observation that users tend to retweet content that is consistent with their world view.  This stems from \emph{selective exposure}, which is a cognitive bias that leads people to avoid the cognitive overload from exposure to opposing views as well as the cognitive dissonance in which people are forced to reconcile between their views and opposing views \cite{morgan2013news}. 

\noindent Concerning media, \newcite{ribeiro2018media} used the Facebook advertising services to infer the ideological leaning of online media based on the political leaning of Facebook users who consumed them. \newcite{an2012visualizing} relied on follow relationships to online media on Twitter to ascertain ideological leaning of media and users based on the similarity between them. \newcite{wong2013quantifying} studied retweet behavior to infer the ideological leanings of online media sources and popular Twitter accounts. \newcite{barbera2015follow} proposed a statistical model based on the follower relationships to media sources and Twitter personalities in order to estimate their ideological leaning.

As for individual users, much recent work focused on stance detection to determine a person's position on a topic including the deduction of political preferences ~\cite{barbera2015birds, barbera2014understanding, borge2015content, cohen2013classifying, colleoni2014echo, conover2011predicting, fowler2011causality, hasan-ng-2014-taking,himelboim2013birds,
magdy2016isisisnotislam, magdy2016failedrevolutions, makazhanov2014predicting, trabelsi2018unsupervised, weber2013secular}. 
User stance classification is aided by the tendency of users to form so-called ``echo chambers'', where they engage with like-minded users \cite{himelboim2013birds,magdy2016isisisnotislam}, and the tendency of users' beliefs to be persistent over time \cite{borge2015content,magdy2016isisisnotislam,pennacchiotti2011machine}.  

Studies have examined the effectiveness of different features for stance detection, including textual features such as word $n$-grams and hashtags, network interactions such as retweeted accounts and mentions, and profile information such as user location
\cite{borge2015content,hasan-ng-2013-stance,magdy2016isisisnotislam,magdy2016failedrevolutions, weber2013secular}.  Network interaction features were shown to yield better results compared to using textual features 
\cite{magdy2016isisisnotislam,wong2013quantifying}.
\newcite{sridhar-etal-2015-joint} leveraged both user interactions and textual information when modeling stance and disagreement, using a probabilistic programming system that allows models to be specified using a declarative language.

\newcite{trabelsi2018unsupervised} described an unsupervised stance detection method that determines the viewpoints of comments and of their authors. It analyzes online forum discussion threads, and therefore assumes a certain structure of the posts.

\noindent It also assumes that users tend to reply to each others' comments when they are in disagreement, whereas we assume the opposite in this paper. Their model leverages the posts' contents, whereas we only use the retweet behavior of users.

Many methods involving supervised learning were proposed for stance detection. Such methods require the availability of an initial set of labeled users, and they use some of the aforementioned features for classification \cite{darwish2017predicting,magdy2016failedrevolutions,pennacchiotti2011democrats}.  Such classification can label users with precision typically ranging between 70\% and 90\% \cite{rao2010classifying,pennacchiotti2011democrats}.  Label propagation is a semi-supervised method that starts with a seed list of labeled users and propagates the labels to other users who are similar based on the accounts they follow or retweet \cite{barbera2015follow,borge2015content,weber2013secular}.  While label propagation may label users with high precision (often above 95\%), it is biased towards users with more extreme views; moreover, careful choice of thresholds is often required, and post-checks are needed to ensure quality.

\newcite{abu-jbara-etal-2013-identifying} and more recently \newcite{darwish2019unsupervisedStance} used unsupervised stance detection, where users are mapped into a lower dimensional space based on user-user similarity, and then clustered to find core sets of users representing different stances. 
This was shown to be highly effective with nearly perfect clustering accuracy for polarizing topics, and it requires no manual labeling of users. Here, we use the same idea, but we combine it with supervised classification based on retweets in order to increase the number of labeled users \cite{darwish2018scotus}.  Other methods for user stance detection include collective classification \cite{duan2012graph}, where users in a network are jointly labeled and classification in a low-dimensional user-space \cite{darwish2017improved}.

As for predicting political leaning or sentiment, this problem was studied previously as a supervised learning problem, where a classifier learns from a set of manually labeled tweets~\cite{pla-hurtado-2014-political,bakliwal-etal-2013-sentiment,bermingham-smeaton-2011-using}. Similarly, \newcite{volkova-etal-2014-inferring} predicted Twitter users' political affiliation (being Republican or Democratic), using their network connections and textual information, relying on user-level annotations.

\section{Data Collection}
\label{sec:data}

We obtained data on eight topics that are considered polarizing in the USA~\cite{darwish2019unsupervisedStance}, shown in Table~\ref{tab:topics}.

\begin{table*}
    \centering
    \tiny
    \begin{tabular}{p{2cm}p{8cm}p{2cm}r}
    \toprule
    \bf Topic & \bf Keywords & \bf Date Range & \bf No. of Tweets \\
    \midrule
    Climate change & \#greendeal, \#environment, \#climate, \#climatechange, \#carbonfootprint, \#climatehoax, \#climategate, \#globalwarming, \#agw, \#renewables  & Feb 25--Mar 4, 2019 & 1,284,902 \\ \hline
    Gun control/rights & \#gun, \#guns, \#weapon, \#2a, \#gunviolence, \#secondamendment, \#shooting, \#massshooting, \#gunrights, \#GunReformNow, \#GunControl, \#NRA & Feb 25--Mar 3, 2019 & 1,782,384 \\ \hline
    Ilhan Omar remarks on Israel lobby & IlhanOmarIsATrojanHorse, \#IStandWithIlhan, \#ilhan, \#Antisemitism, \#IlhanOmar, \#IlhanMN, \#RemoveIlhanOmar, \#ByeIlhan, \#RashidaTlaib, \#AIPAC, \#EverydayIslamophobia, \#Islamophobia, \#ilhan & Mar 1--9, 2019 & 2,556,871 \\ \hline
    Illegal immigration & \#border, \#immigration, \#immigrant, \#borderwall, \#migrant, \#migrants, \#illegal, \#aliens & Feb 25--Mar 4, 2019 & 2,341,316 \\ \hline
    Midterm & midterm, election, elections & Oct 25--27, 2018 & 520,614 \\ \hline
    Racism \& police brutality & \#blacklivesmatter, \#bluelivesmatter, \#KKK, \#racism, \#racist, \#policebrutality, \#excessiveforce, \#StandYourGround, \#ThinBlueLine & Feb 25--Mar 3, 2019 & 2,564,784 \\ \hline
    Kavanaugh Nomination & Kavanaugh, Ford, Supreme, judiciary, Blasey, Grassley, Hatch, Graham, Cornyn, Lee, Cruz, Sasse, Flake, Crapo, Tillis, Kennedy, Feinstein, Leahy, Durbin, Whitehouse, Klobuchar, Coons, Blumenthal, Hirono, Booker, Harris &  Sept. 28-30, 2018 \& Oct. 6-9, 2018 & 2,322,141 \\ \hline
    Vaccination benefits \& dangers & \#antivax, \#vaxxing, \#BigPharma, \#antivaxxers, \#measlesoutbreak, \#Antivacine, \#VaccinesWork, \#vaccine, \#vaccines, \#Antivaccine, \#vaccinestudy, \#antivaxx, \#provaxx, \#VaccinesSaveLives, \#ProVaccine, \#VaxxWoke, \#mykidmychoice & Mar 1--9, 2019 & 301,209 \\
    \bottomrule
    \end{tabular}
    \caption{\label{tab:topics}Polarizing topics used in study.}
\end{table*}

They include a mix of long-standing issues such as racism and gun control, temporal issues such as the nomination of Judge Brett Kavanaugh to the US Supreme Court and Representative Ilhan Omar's polarizing remarks, as well as non-political issues such as the potential dangers of vaccines.  Further, though long-standing issues typically show right--left polarization, stances towards Omar's remarks are not as clear, with divisions on the left as well. 

Since we are interested in US users, we filtered some tweets to retain such by users who have stated that their location was USA.  We used a gazetteer that included words that indicate USA as a country (e.g.,~America, US), as well as state names and their abbreviations (e.g.,~Maryland, MD).

Other data that we used in our experiments is a collection of articles that were cited by users from the tweets collection and that originate from media, whose bias is known, i.e., is discussed on the Media Bias/Fact Check website.

\section{User Stance Detection}
\label{sec:method}

In order to analyze the stance of influencers on a given topic, we first find the stances of Twitter users, and then we project them to the influencers that the users cite. A central (initial) assumption here is that if a user includes a link to some article in their tweet, they are more likely to agree or endorse the article's message. Similarly, when a user retweets a tweet verbatim without adding any comments, they are more likely to agree with that tweet. We label a large number of users with their stance for each topic using a two-step approach, namely \textit{\textbf{projection and clustering}} and \textbf{\textit{supervised classification}}.

For the projection and clustering step, we identify clusters of core vocal users using the unsupervised method described in \cite{darwish2019unsupervisedStance}.  In this step, users are mapped to a lower dimensional space based on their similarity, and then they are clustered.  After performing this unsupervised learning step, we train a supervised classifier using the two largest identified clusters in order to tag many more users.  For that, we use FastText, a deep neural network text classifier, that has been shown to be effective for various text classification tasks \cite{joulin2016bag}.

Once we have expanded our sets of labeled users, we identify influencers that are most closely associated with each group using a modified version of the so-called \emph{valence score}, which varies in value between $-$1 and 1. If an influencer is being cited evenly between the groups, then it would be assigned a valence score close to zero.  Conversely, if one group disproportionately cites an influencer compared to another group, then it would be assigned a score closer to $-$1 or 1.  We perform these steps for each of the given topics, and finally we summarize the stances across all topics.
Below, we explain each of these steps in more detail.

\subsection{Projection and Clustering} 
\begin{figure}
    \centering
    \includegraphics[width=\linewidth]{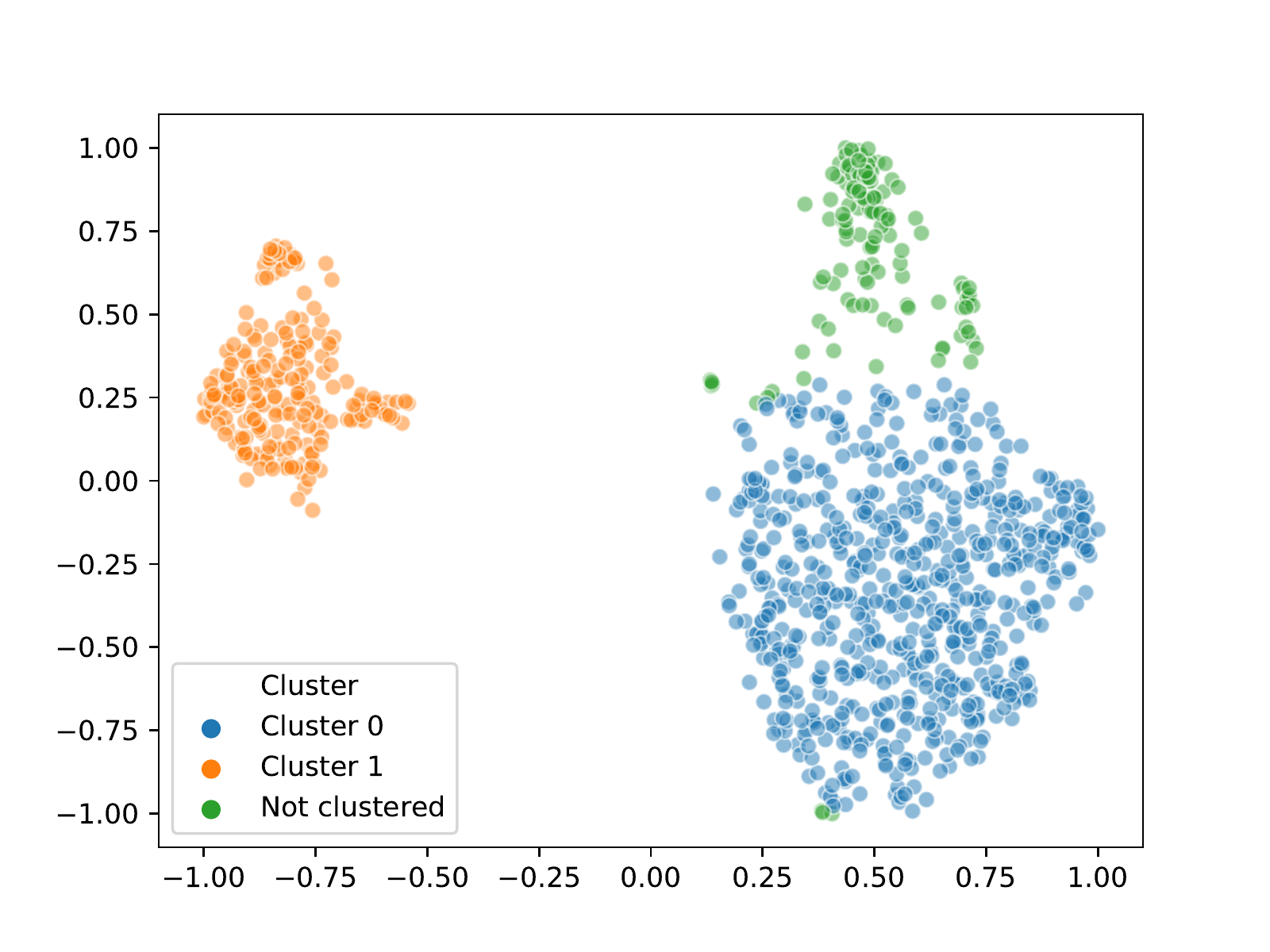}
    \caption{Top active users on the \emph{midterm} topic clustered using UMAP + Mean Shift.}
    \label{fig:umapSample}
\end{figure}

Given the tweets for each topic, we compute the similarity between the top 1,000 most active users. To compute similarity, we construct a vector for each user containing the number of all the accounts that a user has retweeted, and then we compute the pairwise cosine similarity between them.  For example, if user A has only retweeted user B 3 times, user C 5 times and user E 8 times, then user A's vector would be (0, 3, 5, 0, 8, 0, 0, ... 0). Solely using the retweeted accounts as features has been shown to be effective for stance classification \cite{darwish2019unsupervisedStance,magdy2016isisisnotislam}.  Finally, we perform dimensionality reduction and we project the users using Uniform Manifold Approximation and Projection (UMAP).  When performing dimensionality reduction, UMAP places users on a two-dimensional plane such that similar users are placed closer together and dissimilar users are pushed further apart.  Figure~\ref{fig:umapSample} shows the top users for the ``midterm'' topic projected with UMAP onto the 2D plane. After the projection, we use Mean Shift to cluster the users as shown in Figure~\ref{fig:umapSample}. This is the best setup described in \cite{darwish2019unsupervisedStance}.  Clustering high-dimensional data often yields suboptimal results, but can be improved by projecting to a low-dimensional space~\cite{darwish2019unsupervisedStance}.

\subsection{Supervised Classification}
Since unsupervised stance detection is only able to classify the most vocal users, which only constitute a minority of the users, we wanted to assign stance labels to as many additional users as we can. Given the clusters of users that we obtain for each topic, we retain the two largest clusters for each topic, and we assign cluster labels to the users contained therein. Next, we use all the automatically labeled users for each topic to train a supervised classifier using the accounts that each user retweeted as features (same as the features we used to compute user similarity earlier). For classification, we train a FastText model using the default parameters, and then we classify all other users with five or more retweeted accounts, only accepting the classification if FastText was more than 80\% confident (70--90\% yielded nearly identical results).

In order to obtain a rough estimate of the accuracy of the model, we trained FastText using a random 80\% subset of the clustered users for each topic and we tested on the remaining 20\%.  The accuracy was consistently above 95\% for all topics.  This does not mean that this model can predict the stance for all users that accurately --- the clustered users were selected to be the most active ones. Rather, it shows that the classifier can successfully capture what the previous, unsupervised step has already learned. Table~\ref{tab:UsersPerTopic} lists the total number of users who authored the tweets for each topic, the number of users who were automatically clustered using the aforementioned unsupervised clustering technique, and the number of users who were automatically labeled afterwards using supervised classification. Given that we applied unsupervised stance detection to the most active 1,000 users, the majority of the users appeared in the largest two clusters (shown in Table~\ref{tab:UsersPerTopic}).

\begin{table}
    \centering
    \scriptsize
    \begin{tabular}{lrrr}
    \toprule
    \bf Topic & \bf No. of Users & \bf Clustered & \bf Classified \\
     &  & \bf Users & \bf Users \\    \midrule
climate change    & 724,470  & 860 & 5,851 \\
gun control    & 973,206 & 813 & 11,281 \\
Ilhan Omar    & 563,706 & 723 & 25,484 \\
immigration    & 940,840 & 901 & 22,456 \\
midterm elections    & 312,954 & 860 & 12,765 \\
police brutality \& racism    & 1,175,081 & 891 & 18,978 \\
Kavanaugh    & 809,835 & 891 & 10,100 \\
vaccine    & 194,245 & 545 & 556 \\
\bottomrule
    \end{tabular}
    \caption{\label{tab:UsersPerTopic}Users per topic: total number of users, umber of clustered users, and number of automatically labeled users.}
\end{table}

\subsection{Calculating Valence Scores}

Given all the labeled users for each topic, we computed a valence score for each influencer. As mentioned earlier, the valence score ranges between $\left[-1, 1\right ]$, where a value close to 1 implies it is strongly associated with one group of users, $-$1 shows it is strongly associated with the other group of users, and 0 means that it is being shared or cited by both groups.  The original valence score described by \newcite{conover2011political} is calculated as follows:

\vspace{-6pt}
\begin{equation}
    V(u) = 2 \frac{
        \frac{tf(u, C_0)}{total(C_0)}}
        {\frac{tf(u, C_0)}{total(C_0)} + \frac{tf(u, C_1)}{total(C_1)}
    } - 1
    \label{eq:normal_valence}
\end{equation}

\noindent where $tf(u, C_0)$ is the number of times (term frequency) item $u$ is cited by group $C_0$, and $total(C_0)$ is the sum of the term frequencies of all items cited by $C_0$. $tf(u, C_1)$ and $total(C_1)$ are defined in a similar fashion.

\noindent We use the above equation to compute valence scores for the retweeted accounts, but we using a modified version for calculating the score for influencers ($I$):

\begin{equation}
    V(I) = 2 \frac{
        \frac{tf(I, C_0)}{total(C_0)}}
        {\frac{tf(I, C_0)}{total(C_0)} + \frac{tf(I, C_1)}{total(C_1)}
    } - 1
\end{equation}

\noindent where
    
    $tf(I, C_i) = \sum_{a \in I \bigcap C_i} [ln(Cnt(a, C_i)) + 1]$
    
    $total(C_i) = \sum_{I} tf(I, C_i)$\\

In the latter equation, $Cnt(a, C_i)$ is the number of times article $a$ was cited by users from cluster $C_i$.  In essence, we are replacing term frequencies with the natural log of the term frequencies.  We opted to modify the equation in order to tackle the following issue: if users from one of the clusters, say $C_1$, cite only one single article from some media source a large number of times (e.g.,~2,000 times), while users from the other cluster ($C_0$) cite 10 other articles from the same media 50 times each, then using equation \ref{eq:normal_valence} would result in a valence score of $-$0.6. We would then regard the given media as having an opposing stance to the stance of users in $C_0$. Alternatively, using the natural log would lead to a valence score close to 0.88.  Thus, dampening term frequencies using the natural log has the desired effect of balancing between the number of articles being cited by each group and the total number of citations.  We bin the valence scores between $-$1 and 1 into five equal size bands as follows:
\vspace{-6pt}
\begin{equation}
    Cat(V) =
    \begin{cases}
        --, & \text{if } s \in \left [ -1, -0.6 \right )\\
        -, & \text{if } s \in \left [ -0.6, -0.2 \right )\\
        0, & \text{if } s \in \left [ -0.2, 0.2 \right )\\
        +, & \text{if } s \in \left [ 0.2, 0.6 \right )\\
        ++, & \text{if } s \in \left [ 0.6, 1 \right ]
    \end{cases}
    \label{eq:valence_to_cat}
\end{equation}

\section{Characterizing the Influencers}
\label{sec:influencers}

We use valence to characterize the leaning of all cited influencers for each of the topics.  Table~\ref{tab:media_valence_cats} shows the valence categories for the top-cited media sources across all topics. It also shows each media's factuality of reporting, i.e.,~trustworthiness, and bias (ranging from far-left to far-right) as determined by \url{mediaBiasFactCheck.com}. Since the choice of which cluster should be $C_0$ and which would be $C_1$ is arbitrary, we can multiply by $-1$ the valence scores for any topic and the meaning of the results would stay the same. 

We resorted to doing so for some topics in order to align the extreme valence bands across all topics.  Given tweet samples from users in a given cluster for a given topic, labeling that cluster manually was straightforward with almost no ambiguity. Table~\ref{tab:distributionOfMediaAcrossTopics} shows the most frequently cited media source for each topic and for each valence band.  

\begin{table*}[tbh]
    \centering
    \tiny
    \begin{tabular}{|l|c|c||c||c|c|c|c|c|c|c|c|}
\hline
\bf Medium & \bf \begin{turn}{90}factuality\end{turn} & \bf \begin{turn}{90}bias\end{turn} & \bf \begin{turn}{90}Average\end{turn} & \bf \begin{turn}{90}climate change\end{turn} & \bf \begin{turn}{90}gun control\end{turn} & \bf \begin{turn}{90}ilhan\end{turn} & \bf \begin{turn}{90}immigration\end{turn} & \bf \begin{turn}{90}midterm\end{turn} & \bf \begin{turn}{90}police \& racism\end{turn} & \bf \begin{turn}{90}Kavanaugh\end{turn} & \bf \begin{turn}{90}vaccine\end{turn} \\ 
\hline
thehill.com & H & L-C & \pmb{$+$} & 0 & $++$ & $+$ & $+$ & $+$ & $+$ & $++$ & $++$ \\ \hline
theguardian.com & H & L-C & \pmb{$++$} & $++$ & $++$ & $++$ & $++$ & $++$ & $++$ & $++$ & $++$ \\ \hline
washingtonpost.com & H & L-C & \pmb{$++$} & $++$ & $++$ & $++$ & $++$ & $++$ & $++$ & $++$ & $++$ \\ \hline
breitbart.com & VL & Far R & \pmb{$--$} & $--$ & $--$ & $--$ & $--$ & $--$ & $--$ & $--$ & $--$ \\ \hline
foxnews.com & M & R & \pmb{$--$} & $--$ & $--$ & $--$ & $--$ & $--$ & $--$ & $--$ &  \\ \hline
nytimes.com & H & L-C & \pmb{$++$} & $+$ & $++$ & $+$ & $+$ & $+$ & $++$ & $++$ & $++$ \\ \hline
cnn.com & M & L & \pmb{$+$} & $+$ & $++$ & $+$ & $++$ & $+$ & $+$ & $++$ & $+$ \\ \hline
apple.news &  &  & \pmb{$+$} & 0 & 0 & $+$ & 0 & 0 & $+$ & $+$ & $++$ \\ \hline
dailycaller.com & M & R & \pmb{$--$} & $--$ & $--$ & $--$ & $--$ & $--$ & $--$ & $--$ &  \\ \hline
rawstory.com & M & L & \pmb{$++$} & $++$ & $++$ & $++$ & $++$ & $++$ & $++$ & $++$ & $++$ \\ \hline
huffingtonpost.com & H & L & \pmb{$++$} & $++$ & $++$ & $++$ & $++$ & $+$ & $++$ & $++$ & $++$ \\ \hline
truepundit.com & L &  & \pmb{$--$} & $--$ & $--$ & $--$ & $--$ & $--$ & $--$ & $--$ & $--$ \\ \hline
nbcnews.com & H & L-C & \pmb{$+$} & $--$ & $++$ & $+$ & $++$ & $+$ & $+$ & $++$ & $++$ \\ \hline
westernjournal.com & M & R & \pmb{$--$} & $--$ & $--$ & $--$ & $--$ & $--$ & $--$ & $--$ &  \\ \hline
reuters.com & VH & C & \pmb{$+$} & $+$ & $++$ & $++$ & $+$ & $+$ & $+$ & $+$ & $++$ \\ \hline
washingtonexaminer.com & H & R & \pmb{$--$} & $--$ & $--$ & $--$ & $--$ & 0 & $--$ & $--$ &  \\ \hline
thegatewaypundit.com & VL & Far R & \pmb{$--$} & $--$ & $--$ & $--$ & $--$ & $--$ & $--$ & $--$ &  \\ \hline
politico.com & H & L-C & \pmb{$+$} & $+$ & $+$ & $+$ & $+$ & $++$ & $+$ & $+$ & $++$ \\ \hline
npr.org & VH & L-C & \pmb{$+$} & 0 & $++$ & $++$ & $++$ & 0 & $++$ & $++$ & $++$ \\ \hline
townhall.com & M & R & \pmb{$--$} & $--$ & $--$ & $--$ & $--$ & $--$ & $--$ & $--$ & $--$ \\ \hline
msn.com & H & L-C & \pmb{$+$} & $+$ & $+$ & $+$ & 0 & $++$ & 0 & $++$ & 0 \\ \hline
nypost.com & M & R-C & \pmb{$-$} & $--$ & 0 & $-$ & $-$ & $+$ & $--$ & $-$ &  \\ \hline
vox.com & H & L & \pmb{$++$} & $++$ & $++$ & $++$ & $++$ & $++$ & $+$ & $++$ & $++$ \\ \hline
thedailybeast.com & H & L & \pmb{$++$} & $++$ & $++$ & $+$ & $++$ & $++$ & $+$ & $++$ & $++$ \\ \hline
bbc.com & H & L-C & \pmb{$+$} & $+$ & $+$ & $++$ & $++$ & 0 & $+$ & $+$ & $++$ \\
\hline
independent.co.uk & H & L-C & \pmb{$++$} & $++$ & $+$ & $++$ & $++$ & $++$ & $+$ & $++$ & $++$ \\ \hline
ilovemyfreedom.org & VL & Far R & \pmb{$--$} & $--$ & $--$ & $--$ & $--$ & $--$ & $--$ & $--$ &  \\ \hline
thinkprogress.org & M & L & \pmb{$++$} & $++$ & $++$ & $++$ & $++$ & $++$ & $++$ & $++$ & $++$ \\ \hline
dailywire.com & M & R & \pmb{$--$} & $--$ & $--$ & $--$ & $--$ & $--$ & $--$ & $--$ & $++$ \\ \hline
pscp.tv &  &  & \pmb{$-$} & $--$ & $--$ & $--$ & 0 & $--$ & 0 & $-$ &  \\ \hline
dailymail.co.uk & VL & R & \pmb{$-$} & $-$ & 0 & $-$ & $-$ & $-$ & $-$ & $--$ & $--$ \\ \hline
msnbc.com & M & L & \pmb{$++$} & $++$ & $++$ & $++$ & $++$ & $+$ & $++$ & $++$ &  \\ \hline
dailykos.com & M & L & \pmb{$++$} & $++$ & $++$ & $++$ & $++$ & $+$ & $++$ & $++$ &  \\ \hline
bloomberg.com & H & L-C & \pmb{$+$} & $+$ & $++$ & 0 & $++$ & $+$ & 0 & $+$ & $++$ \\ \hline
usatoday.com & H & L-C & \pmb{$+$} & $+$ & $+$ & 0 & $+$ & $++$ & $+$ & 0 & $+$ \\ \hline
    \end{tabular}
\caption{\label{tab:media_valence_cats}Media valence categories for each topic with included average column. Plus ($+$) and minus ($-$) signify left or right leaning, respectively. Factuality: Very High (VH), High (H), Mixed (M), Low (L), Very Low (VL). Bias: Left (L), Left-Center (L-C), Center (C), Right-Center (R-C), Right (R), Far Right (Far R). Blank cells mean that we did not have information.}
\end{table*}

\begin{table*}
    \centering
    \tiny
    \setlength\tabcolsep{1.5pt}
    \begin{tabular}{@{}p{2.5cm}|c|c|p{2.5cm}|c|c|p{2.5cm}|c|c|p{2.5cm}|c|c@{}}
\multicolumn{3}{c|}{\bf climate change} & \multicolumn{3}{c|}{\bf gun control} & \multicolumn{3}{c|}{\bf Ilhan Omar} & \multicolumn{3}{c}{\bf immigration} \\ \hline
theguardian.com    & H  & L-C & thehill.com    & H  & L-C & washingtonpost.com    & H  & L-C & theguardian.com    & H  & L-C \\
washingtonpost.com    & H  & L-C & cnn.com    & M  & L & theguardian.com    & H  & L-C & washingtonpost.com    & H  & L-C \\
independent.co.uk    & H  & L-C & nytimes.com    & H  & L-C & mondoweiss.net    & H  & L & cnn.com    & M  & L \\
wef.ch    &    &   & npr.org    & VH  & L-C & thinkprogress.org    & M  & L & huffingtonpost.com    & H  & L \\
vox.com    & H  & L & washingtonpost.com    & H  & L-C & haaretz.com    & H  & L-C & npr.org    & VH  & L-C \\ \hline
nytimes.com    & H  & L-C & politico.com    & H  & L-C & nytimes.com    & H  & L-C & thehill.com    & H  & L-C \\
bbc.com    & H  & L-C & usatoday.com    & H  & L-C & thehill.com    & H  & L-C & nytimes.com    & H  & L-C \\
cnn.com    & M  & L & msn.com    & H  & L-C & politico.com    & H  & L-C & reuters.com    & VH  & C \\
reuters.com    & VH  & C & bbc.com    & H  & L-C & cnn.com    & M  & L & politico.com    & H  & L-C \\
bloomberg.com    & H  & L-C & cnbc.com    & H  & L-C & apple.news    &    &   & usatoday.com    & H  & L-C \\ \hline
thehill.com    & H  & L-C & apple.news    &    &   & mediaite.com    & H  & L & apple.news    &    &   \\
apple.news    &    &   & sun-sentinel.com    & H  & R-C & usatoday.com    & H  & L-C & msn.com    & H  & L-C \\
npr.org    & VH  & L-C & nypost.com    & M  & R-C & yahoo.com    & M  & L-C & pscp.tv    &    &   \\
seattletimes.com    & H  & L-C & dailymail.co.uk    & VL  & R & timesofisrael.com    & H  & L-C & whitehouse.gov    & M  & R \\
newsweek.com    & M  & L & mailchi.mp    &    &   & theatlantic.com    & H  & L-C & texastribune.org    & H  & C \\ \hline
change.org    & H  & L & washingtontimes.com    & H  & R-C & nypost.com    & M  & R-C & dailymail.co.uk    & VL  & R \\
latimes.com    & H  & L-C & breaking911.com    & VL  &   & jpost.com    & H  & C & nypost.com    & M  & R-C \\
dailymail.co.uk    & VL  & R & chicagotribune.com    & H  & R-C & dailymail.co.uk    & VL  & R & zerohedge.com    & M  &   \\
climatechangedispatch.com    &    &   & rt.com    & M  & R-C & algemeiner.com    & H  & R-C & ir.shareaholic.com    &    &   \\
cnbc.com    & H  & L-C & forbes.com    & M  & R-C & startribune.com    & H  & L-C & breaking911.com    & VL  &   \\ \hline
forbes.com    & M  & R-C & breitbart.com    & VL  & Far R & foxnews.com    & M  & R & breitbart.com    & VL  & Far R \\
breitbart.com    & VL  & Far R & foxnews.com    & M  & R & breitbart.com    & VL  & Far R & illegalaliencrimereport.com    &    &   \\
dailycaller.com    & M  & R & ammoland.com    & H  & R & townhall.com    & M  & R & washingtonexaminer.com    & H  & R \\
tambonthongchai.com    &    &   & dailycaller.com    & M  & R & change.org    & H  & L & foxnews.com    & M  & R \\
wattsupwiththat.com    & L  &   & bearingarms.com    & M  & R & hannity.com    &    &   & westernjournal.com    & M  & R \\ \hline 
\multicolumn{12}{c}{}\\
\multicolumn{3}{c|}{\bf midterm}  & \multicolumn{3}{c|}{\bf police \& racism} & \multicolumn{3}{c|}{\bf Kavanaugh} & \multicolumn{3}{c}{\bf vaccine} \\ \hline
washingtonpost.com    & H  & L-C & washingtonpost.com    & H  & L-C & thehill.com    & H  & L-C & thehill.com    & H  & L-C \\
theguardian.com    & H  & L-C & rawstory.com    & M  & L & washingtonpost.com    & H  & L-C & theguardian.com    & H  & L-C \\
rawstory.com    & M  & L & huffingtonpost.com    & H  & L & cnn.com    & M  & L & washingtonpost.com    & H  & L-C \\
tacticalinvestor.com    &    &   & theguardian.com    & H  & L-C & nytimes.com    & H  & L-C & vaxopedia.org    &    &   \\
vox.com    & H  & L & nytimes.com    & H  & L-C & huffingtonpost.com    & H  & L & nytimes.com    & H  & L-C \\ \hline
thehill.com    & H  & L-C & thehill.com    & H  & L-C & politico.com    & H  & L-C & cnn.com    & M  & L \\
reuters.com    & VH  & C & apple.news    &    &   & apple.news    &    &   & statnews.com    & H  & C \\
nytimes.com    & H  & L-C & cnn.com    & M  & L & yahoo.com    & M  & L-C & latimes.com    & H  & L-C \\
cnn.com    & M  & L & nbcnews.com    & H  & L-C & apnews.com    & VH  & C & cbc.ca    & H  & L-C \\
dailykos.com    & M  & L & thedailybeast.com    & H  & L & latimes.com    & H  & L-C & usatoday.com    & H  & L-C \\ \hline
apple.news    &    &   & msn.com    & H  & L-C & usatoday.com    & H  & L-C & cdc.gov    & VH  &   \\
sagagist.com.ng    &    &   & pscp.tv    &    &   & mediaite.com    & H  & L & medium.com    & M  & L-C \\
bbc.com    & H  & L-C & bloomberg.com    & H  & L-C & theweek.com    & H  & L-C & newsroom.fb.com    &    &   \\
alzwaaj.com    &    &   & politics.theonion.com    &    &   & lawandcrime.com    &    &   & help.senate.gov    &    &   \\
washingtonexaminer.com    & H  & R & rollcall.com    & VH  & C & cnbc.com    & H  & L-C & msn.com    & H  & L-C \\ \hline
dailymail.co.uk    & VL  & R & mediaite.com    & H  & L & pscp.tv    &    &   & change.org    & H  & L \\
pbs.org    & H  & L-C & dailymail.co.uk    & VL  & R & nypost.com    & M  & R-C & fda.gov    &    &   \\
zerohedge.com    & M  &   & news.sky.com    & H  & L-C & ir.shareaholic.com    &    &   & variety.com    &    &   \\
ajc.com    & H  & L-C & newsone.com    & H  & L-C & rollcall.com    & VH  & C &    &   \\
veritablenouvelordre.forumcanada.org    &    &   & aol.com    & H  & L-C & c-span.org    & VH  & C &    &   \\ \hline
breitbart.com    & VL  & Far R & breitbart.com    & VL  & Far R & foxnews.com    & M  & R & ncbi.nlm.nih.gov    & VH  &   \\
foxnews.com    & M  & R & defensemaven.io    &    &   & truepundit.com    & L  &   & vaccineimpact.com    &    &   \\
dailycaller.com    & M  & R & foxnews.com    & M  & R & dailycaller.com    & M  & R & naturalnews.com    & M  &   \\
ilovemyfreedom.org    & VL  & Far R & thegatewaypundit.com    & VL  & Far R & breitbart.com    & VL  & Far R & vaccines.me    &    &   \\
westernjournal.com    & M  & R & nypost.com    & M  & R-C & thegatewaypundit.com    & VL  & Far R & thevaccinereaction.org    &    &   \\ \hline
    \end{tabular}
    \caption{\label{tab:distributionOfMediaAcrossTopics}Top 5 websites per valence category for each topic.}
\end{table*}

\begin{figure}
    \centering
    \includegraphics[width=\linewidth]{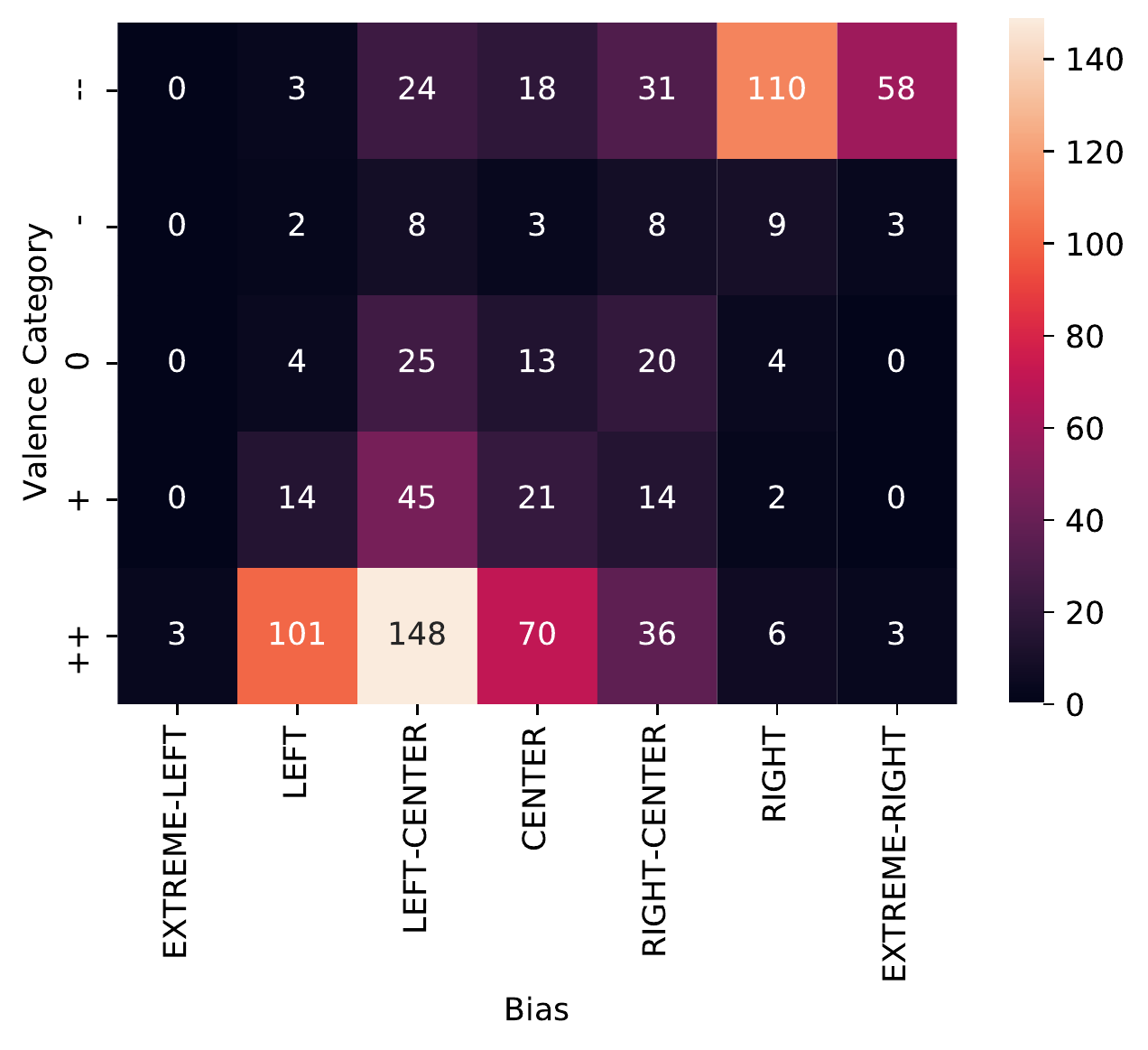}
    \caption{Valence category vs. bias: number of media.}
    \label{fig:valence_bias_corr}
\end{figure}

Of the 5,406 unique media sources that have been cited in tweets across all
topics, 806 have known political bias from \url{mediaBiasFactCheck.com}.
Figure~\ref{fig:valence_bias_corr} shows the confusion matrix between our
valence categories and the goold labels from \url{mediaBiasFactCheck.com}.

We notice that many of the media that have a negative valence score (categories $-$ and $--$) are classified on the right side of the political spectrum by \url{mediaBiasFactCheck.com}, while most media with positive scores (categories $+$ and $++$) are classified as slightly left-leaning. Although there are almost no extreme-left cases, there is a correlation between bias and our valence score.  \url{mediaBiasFactCheck.com} seems to rarely categorize media sources as ``extreme-left''. This could be a reflection of reality or it might imply that \url{mediaBiasFactCheck.com} has an inherent bias.

We also computed the valence scores for the top-200 retweeted accounts, and we assigned each account a valence category based on the score. Independently, we asked a person who is well-versed with US politics to label all the accounts as left, center, or right. When labeling accounts, right-leaning include those expressing support for Trump, the Republican party, and gun rights, opposition to abortion, and disdain for Democrats. 

As for left-leaning accounts, they include those attacking Trump and the Republicans, and expressing support for the Democratic party and for Liberal social positions. If the retweeted account happens to be a media source, we used \url{mediaBiasFactCheck.com}.  Table~\ref{tab:user_valence_cats} compares the per-topic valence for each
retweeted account along with the average category and the true label. 

\begin{table*}
    \centering
    \tiny
    \begin{tabular}{|l||c||c||c|c|c|c|c|c|c|c|}
\hline
\bf Account & \bf \begin{turn}{90}Truth\end{turn} & \bf \begin{turn}{90}Average\end{turn} & \bf \begin{turn}{90}climate change\end{turn} & \bf \begin{turn}{90}gun control\end{turn} & \bf \begin{turn}{90}ilhan\end{turn} & \bf \begin{turn}{90}immigration\end{turn} & \bf \begin{turn}{90}midterm\end{turn} & \bf \begin{turn}{90}police \& racism\end{turn} & \bf \begin{turn}{90}Kavanaugh\end{turn} & \bf \begin{turn}{90}vaccine\end{turn} \\ \hline
realdonaldtrump & R & \pmb{$--$} &  & $0$ & $--$ & $--$ & $--$ & $--$ & $--$ & $--$ \\ \hline
charliekirk11 & R & \pmb{$--$} &  & $--$ & $--$ & $--$ & $--$ & $--$ & $--$ &  \\ \hline
kylegriffin1 & L & \pmb{$++$} & $++$ & $++$ &  & $++$ & $++$ & $++$ & $++$ & $++$ \\ \hline
dbongino & R & \pmb{$--$} & $--$ & $--$ & $--$ & $--$ & $--$ & $--$ & $--$ &  \\ \hline
kamalaharris & L & \pmb{$++$} & $++$ & $++$ &  & $++$ & $++$ & $++$ & $++$ &  \\ \hline
mitchellvii & R & \pmb{$--$} & $--$ & $--$ & $--$ & $--$ & $--$ & $--$ & $--$ &  \\ \hline
realsaavedra & R & \pmb{$--$} & $--$ & $--$ & $--$ & $--$ &  & $--$ & $--$ &  \\ \hline
krassenstein & L & \pmb{$++$} &  & $++$ & $++$ & $++$ & $++$ & $++$ & $++$ & $++$ \\ \hline
realjack & R & \pmb{$--$} & $--$ & $--$ & $--$ & $--$ & $--$ & $--$ & $--$ & $--$ \\ \hline
nbcnews & L & \pmb{$++$} & $++$ & $++$ & $+$ & $++$ & $++$ & $++$ & $++$ & $++$ \\ \hline
education4libs & R & \pmb{$--$} & $--$ & $--$ & $--$ & $--$ & $--$ & $--$ & $--$ &  \\ \hline
nra & R & \pmb{$--$} &  & $--$ &  & $--$ & $--$ &  & $--$ &  \\ \hline
donaldjtrumpjr & R & \pmb{$--$} &  & $--$ &  & $--$ & $--$ & $--$ & $--$ &  \\ \hline
shannonrwatts & L & \pmb{$++$} &  & $++$ & $++$ &  & $++$ & $++$ & $++$ &  \\ \hline
thehill & L & \pmb{$++$} & $++$ & $++$ & $+$ & $++$ & $+$ & $+$ & $++$ & $++$ \\ \hline
realjameswoods & R & \pmb{$--$} &  & $--$ & $--$ & $--$ & $--$ & $--$ & $--$ &  \\ \hline
gopchairwoman & R & \pmb{$--$} &  &  & $--$ & $--$ & $--$ &  & $--$ &  \\ \hline
jackposobiec & R & \pmb{$--$} & $--$ & $--$ & $--$ & $--$ & $--$ & $--$ & $--$ &  \\ \hline
funder & L & \pmb{$++$} & $++$ & $++$ & $++$ & $++$ & $++$ & $++$ & $++$ &  \\ \hline
cnn & L & \pmb{$++$} & $++$ & $++$ & $++$ & $++$ & $0$ & $++$ & $++$ & $++$ \\ \hline
ajplus & L & \pmb{$++$} & $++$ & $++$ & $++$ & $++$ & $++$ & $++$ & $0$ & $++$ \\ \hline
rashidatlaib & L & \pmb{$++$} &  & $++$ & $++$ & $++$ &  & $++$ & $+$ &  \\ \hline
stevescalise & R & \pmb{$--$} &  & $--$ &  & $--$ &  &  & $--$ &  \\ \hline
jordan\_sather\_ & ? & \pmb{$--$} & $--$ &  &  & $--$ & $--$ &  & $--$ & $--$ \\ \hline
aoc & L & \pmb{$++$} & $++$ &  & $++$ & $++$ &  & $++$ &  &  \\ \hline
    \end{tabular}
\caption{\label{tab:user_valence_cats}User valence categories for each topic, preceded by an average
column, and a ground truth label. When a cell is blank, there is \kareem{insufficient} data for
that particular topic. 
}
\end{table*}
\begin{figure}
    \centering
    \scriptsize
    \includegraphics[width=0.95\linewidth]{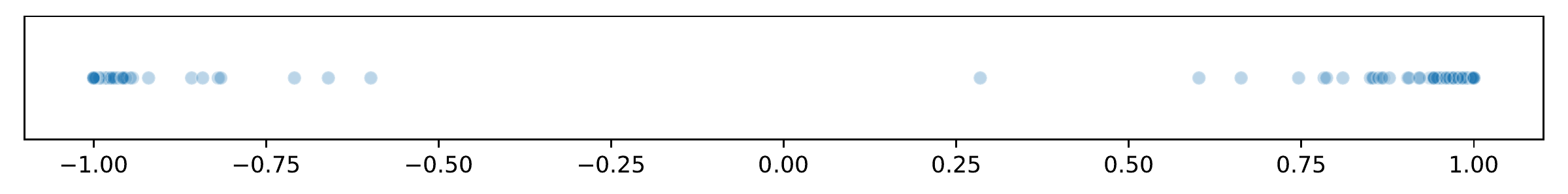}
    \caption{The top-200 retweeted accounts, projected on a number line according to their average valence.}
    \label{fig:retweeted_accounts_valences}
\end{figure}

It is noteworthy that all top-200 retweeted accounts have extreme valence categories on average across all topics.  Their average valence scores, with one exception, appear between $-$0.6 and $-$1.00 for right, and between 0.6 and 1 for left (see Figure~\ref{fig:retweeted_accounts_valences}).

Of those manually and independently tagged accounts, all that were tagged as left-leaning have a strong positive valence score and all that were tagged as right-leaning have a strong negative valence score. Only two accounts were manually labeled as \emph{center}, namely Reuters and CSPAN, which is a US channel that broadcasts Federal Government proceedings, and they had valence scores of 0.55 and 0.28, respectively. Though their absolute values are lower than those of all other sources, they are mapped to the $+$ valence category.

Table~\ref{tab:media_valence_cats} summarizes the valence scores for the media across all topics.  Table~\ref{tab:distributionOfMediaAcrossTopics} lists the most cited media sources for each topic and for each of the five valence bands. The order of the bands from top to bottom is: $++$, $+$, $0$, $-$ and $--$.  The table also includes the credibility and the political leaning tags from \url{mediaBiasFactCheck.com}. The key observations from the table as follows:
\vspace{-6pt}
\begin{enumerate}[wide, labelwidth=!]
    \item Most right-leaning media appear overwhelmingly in the $-$ and $--$ valence categories. Conversely, left-leaning media appear in all valence categories, except for the $--$ category. This implies that left-leaning users cite right-leaning media sparingly. We looked at some instances where right-leaning users cited left-leaning media, and we found that in many cases the cited articles reinforced a right-leaning viewpoint. For example, right-leaning users shared a video from \url{thehill.com}, a left-center site, 2,398 times for the \emph{police racism} topic.  The video  defended Trump against charges of racism by Lynne Patton, a long-time African-American associate of Trump.
\vspace{-6pt}
    \item Most right-leaning sources in the $--$ category have mixed, low, or very low factuality. Conversely, most left-leaning sites appearing in the $-$ valence category have high or very high factuality. Similarly for the vaccine topic, where high credibility sources, such as \url{fda.gov} and \url{nih.gov}, are frequently cited by anti-vaccine users, mostly to support their beliefs.
\vspace{-6pt}
    \item The placements of sources in different categories are relatively stable across topics. For example, \url{washingtonPost.com} and \url{theguardian.com} exclusively appear in the $++$ category, while \url{breitbart.com} and \url{foxnews.com} consistently appear in the $--$ category.
\end{enumerate}

\section{Predicting Media Bias}
\label{sec:media:bias}

Given the stances of users on the aforementioned eight topics, we leverage this information to predict media bias. Specifically, we describe in this section how we make use of the valence scores, as well as other features, namely graph and contextualized text embeddings, to train supervised classifiers for this purpose.
\vspace{-4pt}
\paragraph{Valence Scores.}
We use valence scores in two ways.  First, we average the corresponding valence across the different polarizing topics to obtain an average valence score for a given target news medium.  This is an unsupervised method for computing polarity.  Second, we train a Logistic Regression classifier that uses the calculated valence scores as features and annotations from \url{mediaBiasFactCheck.com} as gold target labels in order to predict the general political leaning of a target news medium.
We merged ``left'' and ``extreme left'',
and similarly we merged ``right'' and ``extreme right''.  We discarded media labeled as being ``left-center'' and ``right-center''.
Each news medium was represented by an 8-dimensional vector containing the valence scores for the above topics.
In the experiments, we used the lbfgs solver and $C = 0.1$.
We used two measures to evaluate its performance, namely accuracy and mean absolute error (MAE). 
The latter is calculated by considering the different classes as ordered and equally distant from each other, i.e.,~if the model predicts \emph{right} and the true label is \emph{left},  this amounts to an error equal to 2. 

The results are shown in Table~\ref{tab:log_reg_maj_scores}, where we can see that using the average valence score yields 68.0\% accuracy (0.330 MAE) compared to 75.2\% accuracy (0.278 MAE) when using the eight individual valence scores as features.

\begin{table}
    \centering
    \scriptsize
\begin{tabular}{@{}lcccc@{}}
\toprule
& \multicolumn{2}{c}{\bf  No Valence} & \multicolumn{2}{c}{\bf  With Valence} \\
& \bf Acc & \bf MAE & \bf Acc & \bf MAE \\
\midrule
Baseline 1 (majority class) & 43.3 & .856 & 43.3 & .856 \\
Baseline 2 (average valence) & -- & -- & 68.0 & .330 \\
Valence scores & -- & -- & 75.2 & .278 \\
\midrule
BERT (article title) & 60.6 & .539 & 78.3 & .264 \\
BERT (article content) & 61.1 & .526 & 79.2 & .255 \\
BERT (title+content) & 62.2 & .510 & 80.8 & .228 \\
\midrule
BERT(Tweet) & 64.0 & .485 & 73.6 & .302 \\
\midrule
GraphEmbM & 63.5 & .468 & 69.1 & .380 \\
GraphEmbH & 66.9 & .425 & 71.8 & .347 \\
GraphEmbM+H & 68.0 & .400 & 79.0 & .251 \\
\midrule
GraphEmbM+H+BERT (tweet) & 72.5 & .358 & 80.5 & .230 \\
GraphEmbM+H+BERT (tweet, content) & 76.1 & .311 & 81.2 & .221 \\
GraphM+H+BERT (tweet, title, content) & \bf 78.1 & \bf .284 & \bf 82.6 & \bf .206 \\
\bottomrule
\end{tabular}
    \caption{\label{tab:log_reg_maj_scores}Predicting media bias.}
\end{table}
\vspace{-4pt}

\paragraph{Graph embeddings.} 
We further use graph embeddings, generated by building a User-to-Hashtag graph (U2H) and a User-to-Mention (U2M) graph and then running node2vec on both \cite{CoNLL2019:troll:roles}, producing two types of graph embeddings. When using graph embeddings, we got worse results compared to our previous setup with valence scores (see Table~\ref{tab:log_reg_maj_scores}). However, when we combine them with the valence scores, we observe a sizable boost in performance, up to 11\% absolute. 
\vspace{-4pt}
\paragraph{Tweets.} We also experimented with BERT-base.
We used the text of the tweets that cite the media we are classifying.  For classification, we fed BERT representations of tweets to a dense layer with softmax output to fine-tune it with the textual contents of the tweets.
We trained at the tweet level, and we averaged the scores (from softmax) for all tweets from the same news medium to obtain an overall label for that news medium. 
The accuracy is much lower than for the valence scores: 64.0\% accuracy vs. 75.2\% for supervised and 68.0\% for unsupervised. 
\vspace{-4pt}
\paragraph{Article titles and text.} Using the BERT setup for \textbf{Tweets}, we used the titles and the full text of up to 100 articles from each of the target media.  When using the full text of articles, we balanced the number of articles per news medium.
We trained two separate BERT models, one on the titles and another one on the full text (content).  Both models did worse than using valence alone, but the combination improved over valence only.
\vspace{-4pt}
\paragraph{System Combination.}  We combined different setups including using all the aforementioned models in combination.  Using graph embeddings (GraphH + GraphM) with BERT embeddings (Tweet+Title+Content) and valence yielded the best results with accuracy of 82.6\%  and MAE of .206.  If we remove valence from the combination, the accuracy drops by 4.5\% while MAE jumps by .078, absolute.  This suggests that valence is a very effective feature that captures important information, complementary to what can be modeled using graph and contextualized text embeddings.

\section{Conclusion and Future Work}
\label{sec:conclusion}

We have presented a method for predicting the general political leaning of media sources and popular Twitter users, 
as well as their stances on specific polarizing topics. Our method uses retweeted accounts, and a combination of dimensionality reduction and clustering algorithms, namely UMAP and Mean Shift, in order to produce sets of users that have opposing opinions on specific topics.  Next, we expand the discovered sets using supervised learning that is trained on the automatically discovered user clusters. We are able to automatically tag large sets of users according to their stance of preset topics. Users' stances are then projected to the influencers that are being cited in the tweets for each of the topics using the so-called \emph{valence score}. The projection allows us to tag a large number of influencers with their stances on specific issues and with their political leaning in general (i.e.,~\emph{left} vs. \emph{right}) with high accuracy and with minimal human effort.
The main advantage of our method is that it does not require manual labeling of entity stances, which requires both
topical expertise and time. We also investigated the quality of the valence features, and we found that valence scores help to predict media bias with high accuracy.

In future work, we plan to increase the number of topics that we use to characterize media.  
Ideally, we would like to automatically identify such polarizing topics.  Doing so would enable us to easily retarget this work to new countries and languages.

\section*{Acknowledgments}
This research is part of the Tanbih project\footnote{\url{http://tanbih.qcri.org/}}, which aims to limit the effect of ``fake news,'' propaganda and media bias by making users aware of what they are reading.

\bibliography{acl2020}
\bibliographystyle{acl_natbib}

\end{document}